# Bounded Budget Connection (BBC) Games or How to make friends and influence people, on a budget

Nikolaos Laoutaris [*]    Laura J. Poplawski [†‖]    Rajmohan Rajaraman [‡‖]

Ravi Sundaram [§‖]    Shang-Hua Teng [¶]

## Abstract

Motivated by applications in social networks, peer-to-peer and overlay networks, we define and study the *Bounded Budget Connection* (BBC) game - we have a collection of $n$ players or nodes each of whom has a budget for purchasing links; each link has a cost as well as a length and each node has a set of preference weights for each of the remaining nodes; the objective of each node is to use its budget to buy a set of outgoing links so as to minimize its sum of preference-weighted distances to the remaining nodes.

We study the structural and complexity-theoretic properties of pure Nash equilibria in BBC games. We show that determining the existence of a pure Nash equilibrium in general BBC games is NP-hard. We counterbalance this result by considering a natural variant, fractional BBC games - where it is permitted to buy fractions of links - and show that a pure Nash equilibrium always exists in such games.

A major focus is the study of $(n,k)$-uniform BBC games - those in which all link costs, link lengths and preference weights are equal (to 1) and all budgets are equal (to $k$). We show that a pure Nash equilibrium or stable graph exists for all $(n,k)$-uniform BBC games and that all stable graphs are essentially fair (i.e. all nodes have similar costs). We provide an explicit construction of a family of stable graphs that spans the spectrum from minimum total social cost to maximum total social cost. To be precise we show that that the price of stability is $\Theta(1)$ and the price of anarchy is $\Omega(\sqrt{\frac{n/k}{\log_k n}})$ and $O(\sqrt{\frac{n}{\log_k n}})$.[1] We also study a special family of *regular* graphs, the class of Abelian Cayley graphs, in which all nodes imitate the "same" buying pattern, and show that if $n$ is sufficiently large no such regular graph can be a pure Nash equilibrium. This has the implication that, as the designer of a P2P or overlay network, one has to give up stability in order to get the simplicity and convenience of regularity.

We analyze best-response walks on the configuration space defined by the uniform game, and show that starting from any initial configuration, strong connectivity is reached within $n^2$ rounds. We demonstrate that convergence to a pure Nash equilibrium is not guaranteed by demonstrating the existence of an explicit loop which also proves that even uniform BBC games are not potential games. Lastly, we extend our results to the case where each node seeks to minimize its maximum distance to the other nodes.

[*]Telefonica Research, nikos@tid.es

[†]ljp@ccs.neu.edu

[‡]rraj@ccs.neu.edu

[§]koods@ccs.neu.edu

[¶]Boston University, steng@cs.bu.edu, supported by NSF ITR CCR-0325630

[‖]Northeastern University, partially supported by NSF grant CCF-0635119

[1]Observe that our bounds for the price of anarchy are essentially tight when $k$ is a constant.

# 1 Introduction

## 1.1 Motivation

You are the campaign manager for a Presidential candidate and it is the start of what will be a long and grueling series of primaries and caucuses to determine your party's nominee. You have a limited budget for the campaign, in terms of money and time. And, you need to understand the (organized as well as informal) networks of connections and influence that exist within the nation to decide how best to allocate your scarce resources so as to have the optimal impact on voters. Many of the players and political operatives you choose to reach out to are not only being courted by other candidates but also have their own ambitions (maybe at the regional or town levels) and agendas. Your actions affect and, in turn, are affected by the actions of the others, voters and candidates, in the race. In the world of ever shifting political loyalties you need to understand the calculus of allegiance: what should you do and who should you ally with so as to effectively counteract and neutralize the strategies of your opponents while maximizing your chances of winning the required votes?

You are the founder of a social networking website, such as a friend finder site or a site where people trade timeshares on vacation homes. Your income from the site depends on how well people are connected to one another. The more easily they can find others to befriend or trade timeshares with, the more money you make. People have natural bounds on their time and cognitive resources and hence are limited in the number of people they can maintain direct ties with (also known as the Dunbar limit in the sociology literature). They must rely on friends of friends, and friends of friends of friends and so on to reach other people in the network. It is no surprise that you are concerned with understanding how the structure of networks generated by individuals expressing their natural preferences and aversions will affect your ability to monetize the network. Could it be possible that left to their own devices people will generate poorly connected networks?

You are designing the next killer application, the next Napster, the next Kazaa, the next big thing in the world of unstructured peer-to-peer file sharing networks or overlay networks. You know that people will hack the open source reference implementation of the client to create nodes that will behave strategically, selecting their first hop neighbors to selfishly optimize their utility. In unstructured P2P file sharing nodes employ scoped flooding or multiple parallel random walks to reach other nodes and thus have to adhere to small out-degrees to prevent clogging. For analogous reasons of scalability, overlay networks too require constraints on the out-degree of nodes so as to reduce the number of links that need monitoring and to reduce the amount of link state information that needs to be disseminated. You know that the success of your killer application depends critically on the connectedness of the network. Every node will independently attempt to minimize its average latency to the subset of nodes of interest, but will this lead to an operating point that is close to the social optimum or can it lead to an anarchic situation characterized by an impoverishment of connectivity?

In this paper we define and study a graph-theoretic game called the Bounded Budget Connection (BBC) game that abstracts each of the three situations above where strategic nodes acting under a *cost budget* form connections (friends) with a view to optimizing their proximity (influence) to the nodes of interest. This is a big problem space that allows for a variety of models to capture different situations. In addition to different notions of connection cost and proximity e.g., fractional and integral, symmetric and asymmetric, uniform and nonuniform, metric spaces, etc., one can also consider a variety of solution and equilibrium concepts (other than Nash equilibria) as well as the dynamics of the resultant complex systems. There are many earlier works that touch on similar issues as detailed in subsection 1.3. We believe the budget constraint is an important real-world restriction and consider our paper to be a preliminary step towards understanding and characterizing the rich and elegant structures that exist in this domain.

## 1.2 Our Results

To capture the above scenarios we posit the following Bounded Budget Connection (BBC) game - we have a collection of $n$ players or nodes each of whom has a budget for purchasing links; each link has a cost as well as a length and each node has a set of preference weights for each of the remaining nodes; the objective of each node is to use its budget to buy a set of outgoing links from itself so as to minimize its sum of preference weighted distances to the remaining nodes.

Our goal in this paper is to study the structural and complexity-theoretic properties of pure Nash equilibria. We first present our results on nonuniform BBC games, the most general kind of BBC games. Nonuniform BBC games are best explained by defining their complement. Uniform BBC games are those in which all link costs are equal, all link lengths are equal, all preference weights are equal and all budgets are equal. Nonuniform BBC games are BBC games which are not uniform BBC games.

- We show that determining the existence of a pure Nash equilibrium in nonuniform BBC games is NP-hard. To be precise, we can show the NP-hardness of determining the existence of a pure Nash equilibrium when link costs, link lengths or preference weights are nonuniform[2]

- We counterbalance this NP-hardness result by showing that in all fractional BBC games a pure Nash equilibrium always exists. A fractional BBC game is one in which it is permitted to buy fractions of links and is a natural "fractionalization" of the integral version. Fractional BBC games capture scenarios in which nodes use links only for a fraction of the time, instead of all the time, e.g., packetized networks.

Next, we present our results on uniform BBC games. We can assume, without loss of generality, that all link weights, link lengths and preference weights are equal to 1 and all budgets are equal to $k$, thus allowing us to talk of $(n,k)$-uniform BBC games.

- We show that a pure Nash equilibrium or stable graph exists for all $(n,k)$-uniform BBC games and that all stable graphs are essentially fair (i.e. all nodes have similar costs). We provide an explicit construction of a family of stable graphs that spans the spectrum from minimum total social cost to maximum total social cost. To be precise we show that that the price of stability is $\Theta(1)$ and the price of anarchy is $\Omega(\sqrt{\frac{n/k}{\log_k n}})$ and $O(\sqrt{\frac{n}{\log_k n}})$. Observe that our bounds for the price of anarchy are essentially tight when $k$ is a constant.

- Inspired by the existence of stable graphs in the uniform case, we next tackle the question of finding stable graphs that are "regular" in a strong sense (to be defined subsequently). Such graphs, if they existed, would have applications to overlay and P2P networks. Unfortunately we are able to show that even the class of Abelian Cayley graphs (a strict superset of "regular" graphs) does not possess a stable graph. In other words, stability and regularity are mutually incompatible. This has the implication that, as the designer of a P2P or overlay network, one has to give up stability in order to get the simplicity and convenience of regularity.

Lastly, we consider the dynamics of best response moves.

- We show that in any $(n,k)$-uniform BBC game, a (suitably defined and entirely natural) best response walk converges to a strongly connected configuration within $n^2$ steps.

- We show that uniform BBC games are not (ordinal) potential games by presenting a loop for best response walks. This serves to underscore the importance of our explicit constructions of stable graphs, as it rules out the possibility of demonstrating existence of Nash equilibria through suitably defined potential functions.

We end by showing that there are analogous results for the case where the cost function is the maximum (instead of sum) of the weighted distances.

---
[2]We believe that this question is not just NP-hard but in fact $\Sigma_2$-complete. Further, we also conjecture that, in fact, pure Nash equilibria do exist in all cases where only the budgets are non-uniform.

## 1.3 Related Work

Notions of group and network formation along with concepts of influence have been investigated by a number of different communities starting with researchers in economics and game theory and followed by work in combinatorial optimization and computer science. The work of [17] modeled and analyzed the stability of networks when nodes themselves choose to form or sever links; their model is different from ours in that they studied different stylized models that included production and allocation functions under the (relatively weak) concept of pairwise stability, along with side payments. [5] study a model of directed network formation where nodes incur costs based on the number of incoming links. In [13], where they defined and studied a similar network creation game, the authors do not have a fixed budget of directed links for the nodes; instead they consider undirected links, and the nodes optimize a cost which is the sum of the number of edges, scaled by a parameter $\alpha > 0$, and the sum of distances to the rest of the nodes. They present several results on the price of anarchy, which is the ratio of the cost of the worst-case Nash equilibrium to the social optimum cost [21]. Further results in this direction are obtained in [1] and [10]. [15] extends this model to the case where each node is only interested in connecting to a subset of the other nodes. [11] is similar to [13] in that they impose a cost for the purchase of a link rather than a fixed budget, however they consider a stochastic model and associated small-world effects. In [24] a variant is studied, in which the nodes are embedded in a metric space and the distance component of the cost is replaced by the stretch with respect to the metric. They obtain tight bounds on the price of anarchy and show that the problem of deciding the existence of pure Nash equilibria is NP-hard. Network formation under the requirement for bilateral consent for building links is studied in [9]. [12] focuses on a similar network creation game restricted to a bipartite graph, with nodes representing buyers and sellers. Our model follows directly in the tradition of [7, 22] where they present experimental studies of network formation games involving non-unit link lengths.

Network formation games have also been studied in the context of Internet inter-domain routing. A coalitional game-theoretic problem modeling of BGP is introduced in [26] and studied further in [23]. A fractional version is studied in [16]. Also related is the work on designing strategy-proof mechanisms for BGP [14] as well as the recent work on strategic network formation through AS-level contracts [4]. [18] consider a contracts-based model of network formation where links do not have predefined costs but are subject to negotiation and nodes attempt to minimize incoming traffic by obtaining compensation in return.

Combinatorial optimization aspects are explored in [19, 20] where the goal is to pick an initial set in a stochastic model with maximal expected influence. This model is extended further in [6] to a competitive setting within the stochastic framework where different players compete (sequentially) to maximize their expected influence.

## 2 Problem Definition

A *Bounded Budget Connection* game (henceforth, a BBC game) is specified by a tuple $\langle V, w, c, \ell, b \rangle$, where $V$ is a set of nodes, $w : V \times V \to \mathbb{Z}$, $c : V \times V \to \mathbb{Z}$, $\ell : V \times V \to \mathbb{Z}$, and $b : V \to \mathbb{Z}$ are functions. For any $u, v \in V$, $w(u, v)$ indicates $u$'s preference for communicating with $v$, $c(u, v)$ denotes the cost of directly linking $u$ to $v$, and $\ell(u, v)$ denotes the length of the link $(u, v)$, if established. For any node $u \in V$, $b(u)$, specifies the budget $u$ has for establishing outgoing directed links: the sum of the costs of the links $u$ establishes should not exceed $b(u)$.

A strategy for node $u$ is a subset $S_u$ of $\{(u, v) : v \in V\}$ such that $\sum_{v:(u,v) \in S_u} c(u, v) \leq b(u)$. Let $S_u$ denote a strategy chosen by node $u$ and let $S = \{S_u : u \in V\}$ denote the collection of strategies. The network formed by $S$ is simply the directed graph $G(S) = (V, E)$ where $E = \bigcup_u S_u$. The utility of a node $u$ in $G(S)$ is given by $-\sum_v w(u, v) d(u, v)$, where $d(u, v)$ is the shortest path from $u$ to $v$ in $G(S)$ according to the lengths given by $\ell$. For convenience, we assume that if no path exists in $G(S)$ from $u$ to $v$, then $d(u, v)$ is given by some large integer $M \gg n \max_{u,v} \ell(u, v)$; we refer to $M$ as the *disconnection penalty*.

Following the standard game-theoretic terminology, we say that a strategy selection $S = \{S_u : u \in V\}$ is stable if it is a pure Nash equilibrium for the BBC game; in particular, for each $u$, $S_u$ is an optimal strategy for $u$ assuming that the strategy for every $v \neq u$ is fixed as in $S$.

A major focus of our work is on *uniform games*, in which (a) $c(u,v)$ is identical for all $u, v$; (b) $w(u,v)$ is identical for all $u, v$; (c) $\ell(u,v)$ is identical for all $u, v$; and (d) $b(u)$ is identical for all $u \in V$. In a uniform game, we may assume without loss of generality that $c(u,v) = w(u,v) = \ell(u,v) = 1$ for all $u,v$, and $b(u) = k$, for all $u \in V$, for some integer $k$. We refer to the preceding uniform game as an $(n,k)$-uniform game where $n = |V|$. We refer to BBC games that are not uniform as *non-uniform* games.

## 3 Nonuniform Games

In this section we show there exist instances of non-uniform BBC games that do not have a pure Nash equilbrium. Furthermore, we prove that it is NP-hard to determine whether a given instance of a non-uniform BBC game has a pure Nash equilibrium. This motivates us to consider a natural variant of BBC games, which we call *fractional BBC games*, in which each node can select *fractions* of links, whose total cost is within the node budget. We show that pure Nash equilibria always exist for fractional non-uniform BBC games.

### 3.1 Nonexistence of pure Nash equilibria and NP-hardness

**Theorem 1.** *For any $n \geq 11$, $k \geq 1$, there exists a nonuniform BBC game with $n$ nodes, nonuniform preferences, uniform link costs, uniform link lengths, and a uniform budget of $k$ for every node, such that the game has no pure Nash equilibrium.*

*Proof.* We first construct a BBC game $G$ with $n = 11$, $k = 1$, uniform costs, nonuniform lengths, and nonuniform preferences, such that $G$ has no pure Nash equilibrium. We then show how to drop the nonuniformity in link lengths and also extend the claim to arbitrary values of $n$ and $k$.

The basic idea is to encode the pay-off structure of a "matching pennies" game [25]. To construct such an instance we define a *gadget* (see Figure 1 for an illustration). Our gadget is made out of two *sub-gadgets*, sub-gadget(0) and sub-gadget(1). For $i \in \{0,1\}$, sub-gadget($i$) consists of five nodes: a central one ($iC$), two bottom ones (left, $iLB$ and right, $iRB$), and two top ones (left, $iLT$ and right, $iRT$). We set the length of every link shown in Figure 1 to be 1, while the length of every omitted link is $L$, where $L$ is chosen suitably large. We also have one additional node $X$ (not depicted in the figure) and set the length of $(0LB, X)$, $(0RB, X)$, $(1LB, X)$, and $(1RB, X)$ to be 1 and that of all other links to X to be $L$.

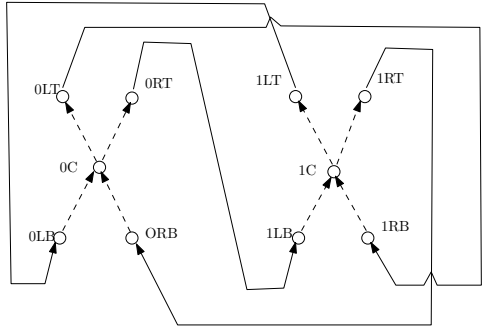

Figure 1: Gadget - consisting of two subgadgets

Having defined the nodes and link lengths, it remains to define the preferences. (Recall that link costs are all uniform and the budget for each node is 1.) For every solid edge $(u,v)$ in Figure 1, we set $w(u,v)$ to be 1. In addition, we set $w(0C, 1C)$, and $w(1C, 0C)$ to be 1. Finally, we set $w(0LB, 0RT)$, $w(0RB, 0LT)$, $w(1LB, 1RT)$, and $w(1RB, 1LT)$ to be 2, and $w(u, X)$ to be 1 for each $u$ in $\{0LB, 0RB, 1LB, 1RB\}$.

We now establish that the instance constructed has no pure Nash equilibrium. Let us consider the case when the action for node $0C$ is link $(0C, 0LT)$. Then, in a stable network, $0RB$ sets its link to $0C$ since it has a higher preference for $0LT$ than $X$, which implies that $1C$ sets its link to $1RT$ and $1RB$ sets its link to $X$. This means that $0C$ does not have a path to $1C$ in the network and will switch its link to $0RT$ to improve its utility. Thus, there is no pure Nash equilibrium that contains the link $(0C, 0LT)$. The other case with the link $(0C, 1LT)$ is symmetric.

We now extend the result to uniform lengths by modifying the preferences. The preferences $w(u, v)$ where $u$ is a top node remain the same. The "switch" from the central node $v$ of a sub-gadget to a top node $u$ of the same sub-gadget can be implemented by setting $w(v, u)$ to be $\zeta > 0$, $w(v, v')$ to be $\xi > 0$, with $\xi < \zeta$, if $v'$ is the central node of the other gadget, and to be 0 otherwise.

Implementing the switch from a bottom node $v$ to either the central node $u$ of the same sub-gadget, or to node $X$ is a little more involved. Let's set $w(v, y) = \alpha$, $w(v, u) = \beta$, and $w(v, v') = \gamma$, where $v'$ denotes $v$'s cross-over node at the top of the same sub-gadget. If $M$ denotes the disconnection penalty we enforce the following inequalities: $\alpha > \gamma$, $\alpha > \beta$, and $\alpha \cdot (M-1) < \beta \cdot (M-1) + \gamma \cdot (M-2)$. The first inequality guarantees that a bottom node will never establish a direct link to its cross-over node at the top of the same sub-gadget. The second one guarantees that if the link from the central node to the cross-over does not exist, then the bottom node will connect to $X$. The last inequality guarantees that if the link from the central node to the cross-over node exists, the bottom node will connect to the central node. The three inequalities can be jointly satisfied by picking positives $\gamma, \epsilon$ such that $\epsilon < \frac{M-2}{M-1} \cdot \gamma$, and setting $\beta = \gamma + \epsilon$ and $\alpha = \beta + \frac{M-2}{M-1} \cdot \gamma - \epsilon$.

This completes the proof for $n = 11$ and $k = 1$. The result easily extends to $n > 11$ or $k \geq 2$ by just forcing all of the remaining links to connect to specific nodes, using appropriate preferences. □

**Theorem 2.** *It is NP-hard to determine whether a given instance of the non-uniform BBC game has a pure Nash equilibrium.*

*Proof.* The proof is by a reduction from 3SAT. Let $\phi$ be a 3SAT formula with $n$ variables and $m$ clauses. We create a non-uniform BBC instance as follows. For each variable $x_i$ in $\phi$, we introduce 3 nodes: a *variable node* $X_i$, and two *truth nodes* $X_{iT}$ and $X_{iF}$. For each clause $c_j$, we introduce a clause node $K_j$ and intermediate nodes $I_{j1}$, $I_{j2}$, and $I_{j3}$, one for each of the three literals in the clause. We also have two additional nodes $S$ and $T$ and a gadget $G$ consisting of the nodes illustrated in Figure 1. Our construction is depicted in Figure 2.

Figure 2: Construction to prove NP-hardness of pure Nash Equilibrium detection

For all $u, v$ in $V$, we set $c(u, v)$ to be 1. The length of every link shown in Figure 2 is 1 and the length of every other link is a large number $L$ greater than the number of nodes; we set the disconnection penalty $M$ to be $nL$. The budget for each node is 1, except for node $S$, which has a budget of $m$.

We now define the preferences and the budgets. Let $V$ denote the set of all nodes. For each $i$, for all $v$, $w(X_{iT}, v)$ and $w(X_{iF}, v)$ are both 0; that is, the truth nodes do not need to communicate with any node in the game. We also set their budgets to 0.

For node $C_i$, we set $w(X_i, v)$ to be 1 for $v \in \{X_{iT}, X_{iF}\}$ and 0 for all other $v$; thus, $X_i$ equally prefers to communicate with $X_{iT}$ and $X_{iF}$ and with no other node. The budget for $X_i$ is set to 1.

Now consider a clause $c_j$. For each intermediate node $I_{jk}$, we set $w(I_{jk}, v)$ to be 1 if $v = X_i$ and 0 otherwise. If the $k$th literal of $c_j$ is $x_i$, then for the intermediate node $I_{jk}$, we set $w(I_{jk}, v)$ to 1 for $v = X_{iT}$ and 0 for all other $v$; else, we set $w(I_{jk}, v)$ to 1 for $v = X_{iF}$ and 0 for all other $v$. For the clause node $K_j$, we set the preferences as follows. If $x_i$ is in clause $K_j$, then $w(K_j, X_{iT})$ is 2; if $\overline{x_i}$ is in clause $K_j$, then $w(K_j, X_{iF})$ is 2; $w(K_j, S)$ is 1 and for all other $v$, $w(K_j, v)$ is 0. The budget for each clause and intermediate node is 1.

We next consider the gadget $G$. The preferences among the top nodes in the gadget are identical to that used in the proof of Theorem 1. We also have $w(0C, 1C)$ and $w(1C, 0C)$ to be $2m - 1$, and $w(0C, v)$ and $w(1C, v)$ to be 2 for each intermediate node $v$. For each bottom node $u$ in $G$, we set $w(u, v)$ to be 3 if $v$ is the cross-over top node in the subgadget, 2 is $v$ is $S$, and 1 for $T$. The budget for each node in the gadget is 1. Finally, we consider nodes $S$ and $T$. For node $S$, $w(S, v) = 1$ if $v$ is a clause node and 0 otherwise. Node $S$ has a budget of $m$ and node $T$ has a budget of 0.

We now show that $\phi$ is satisfiable if and only if the above BBC game has a pure Nash equilibrium. Suppose $\phi$ is satisfiable. Consider a satisfying assignment for $\phi$. If $x_i$ is true, we set the link from $X_i$ to $X_{iT}$; otherwise, we set the link from $X_i$ to $X_{iF}$; in either case, $X_i$ has attained its highest utility possible. The intermediate nodes just link to their respective variable nodes and attain their highest utility. For each clause $c_j$, there exists a literal in the clause, say the $k$th literal, which is satisfied. If the literal equals variable $x_i$, then the intermediate node $I_{jk}$ has a path to $X_{iT}$ through $X_i$. So we set the link from the clause node $C_j$ to $I_{jk}$. A clause node prefers to communicate with three of the truth nodes but can communicate with at most one in any stable network owing to budget constraints. Furthermore, the three-hop path achieved from the clause node to a truth node is the shortest possible, so each clause node has also attained its maximum utility. We finally consider the nodes in the gadget $G$. Each top node sets its link to the only node for which it has a preference. The two central nodes link to node $S$ and achieve their maximum utility possible since they prefer to have paths of length 3 to $m$ of the intermediate nodes over a path of length 3 to the other central node. Each bottom node links to the central node in its subgadget and achieves its maximum utility given the other connections. Thus, the constructed network is stable.

If the BBC game has a pure Nash equilibrium, then each of the central nodes in the gadget $G$ has to link to $S$ since the gadget by itself does not have a pure Nash equilibrium, by the proof of Theorem 1. This occurs only if each of the central nodes has a 3-hop path to at least $m$ intermediate nodes. This in turn implies that each clause node has a link to an intermediate node. A clause node links to an intermediate node only if the intermediate node has a path either to a node $X_{iT}$, where $x_i$ is in the clause, or to a node $X_{iF}$, where $\overline{x_i}$ is in the clause. This is because if no intermediate node for the clause has such a path, then the clause node would link to $S$. This yields the following satisfying assignment for $\phi$: set $x_i$ to true if $X_i$ has a link to $X_{iT}$, and false otherwise.

In the above reduction, the budget function is nonuniform. By using additional nodes, the reduction can be easily adapted to work where the budget of each node is $k$, for $k \geq 2$. □

## 3.2 Fractional BBC games

We consider a natural fractional version of the bounded budget connection game and prove that a pure Nash equilibrium always exists. Following the framework of Section 2, a *fractional BBC* game is specified by a tuple $\langle V, w, c, \ell, b \rangle$, where $V$ is a set of nodes, and $w : V \times V \to \mathbb{Z}$, $c : V \times V \to \mathbb{Z}$, $\ell : V \times V \to \mathbb{Z}$, and $b : V \to \mathbb{Z}$ are functions. As before, $w$, $c$, $\ell$, and $b$ represent the node preference, link cost, link length, and budget functions, respectively. In a fractional game, the strategy space for a node $u$ is $\{a_u : V \to \mathbb{R} \mid \sum_v a_u(v) c(u, v) \leq b(u)\}$. Let $\mathbf{a} = \{a_u : u \in V\}$ denote a collection of strategies, one for each node.

We now define the utility of each node, given an action for each node. In BBC, the cost of a node is the preference-weighted sum of shortest path distances (based on link lengths) to the other nodes in the network determined by the node actions. In the fractional version, the natural equivalent of the shortest path is the cost of a minimum-cost unit flow in a flow network determined by the node actions, where the costs are calculated using the link length function. We also need to account for the case that it may not be possible to send a unit flow from $u$ to $v$ in the flow network; this is analogous to the case in integral BBC games in which there is no path from $u$ to $v$. For a given set of node strategies $\mathbf{a}$, we define the flow network $G(\mathbf{a})$ as follows: for every pair of nodes $u$, $v$, we have two links from $u$ to $v$, one with capacity $a_u(v)$ and length $\ell(i, j)$, and the other with capacity $\infty$ and length $M$, where $M$ is the equivalent of the *disconnection penalty*. The links of the second kind ensure that $G(\mathbf{a})$ always carries a unit flow. Let $\text{cost}_{uv}(\mathbf{a})$ denote the cost of a minimum cost unit flow from $u$ to $v$ in $G(\mathbf{a})$. Then, the utility of node $u$ under $\mathbf{a}$ equals $-\sum_v w(u, v) \text{cost}_{uv}(\mathbf{a})$.

**Theorem 3.** *Every instance of the fractional bounded budget connection game has a pure Nash equilibrium.*

*Proof.* A game has a pure Nash equilibrium if the strategy space of each player is a compact, non-empty, convex space, and the utility function of each player $u$ is continuous on the strategy space of all players and quasi-concave in the strategy space of $u$ [25, Proposition 20.3].

The strategy space of each player $u$ is simply the convex polytope given by $\{\sum_v a_u(v) \leq b(u)\}$. It is clearly compact, non-empty, and convex.

The continuity of the utility function is also clear. It remains to prove that the utility function of each player is quasi-concave in the strategy space of the player. Since the utility function is merely the negative of the cost, we will show that the cost of the min-cost flow is a quasi-convex function.

Consider two strategies $a_u$ and $b_u$ of the player $u$, given fixed strategies of all other players. Let $\mathbf{a}$ denote the strategy-tuple for $a_u$ and $\mathbf{b}$ for $b_u$. Fix a destination $v$. Suppose there exists a unit flow $f_a$ from $u$ to $v$ in $G(\mathbf{a})$ and a unit flow $f_b$ from $u$ to $v$ in $G(\mathbf{b})$. Given any $\lambda \in [0,1]$, consider the strategy $c_u = \lambda a_u + (1-\lambda) b_u$ for player $u$. We define the unit flow $f_c$ from $u$ to $v$ as follows. For any edge $e = (x,y)$, we set $f_c(e) = \lambda f_a(e) + (1-\lambda) f_b(e)$. Since $f_a(e)$ is at most $a_x(y)$ and $f_b(e)$ is at most $b_x(y)$, $f_c(e)$ is at most $\lambda a_x(y) + (1-\lambda) b_x(y) = c_x(y)$. Furthermore, the cost of the flow equals

$$\begin{aligned}
\sum_{(x,y)} f_c((x,y))\ell(x,y) &= \ell(x,y) \sum_{(x,y)} (\lambda f_a((x,y)) + (1-\lambda) f_b((x,y))) \\
&= \lambda \text{cost}_{uv}(\mathbf{a}) + (1-\lambda) \text{cost}_{uv}(\mathbf{b}) \\
&\leq \max\{\text{cost}_{uv}(\mathbf{a}), \text{cost}_{uv}(\mathbf{b})\}.
\end{aligned}$$

Thus, the cost function with respect to one destination is quasi-convex. Since the cost for a player is simply the sum of the weighted costs with respect to all destinations, the quasi-convexity of the cost function and, hence, the quasiconcavity of the utility function follow. This completes the proof of the theorem. □

## 4 Uniform Games

Although non-uniform games lack stability, the simplest version of the framework has many interesting properties. We define a uniform $(n,k)$-BBC game as a game in which all preferences, costs, and lengths are 1, and each node has a budget of $k$ links. In other words, in this graph, all the nodes are equally interested in communicating with all other nodes, any connection can be established for the same cost, and the utility function is calculated using hop counts.

We show that a Nash equilibrium, or stable graph, exists for the uniform $(n,k)$-BBC game with any values of $n$ and $k$ and that all stable graphs are essentially fair (all nodes in a stable graphs have similar cost). We also establish nearly tight bounds on the price of anarchy and price of stability. Although we describe a class of stable graphs, showing that there are multiple Nash equilibria, we show that no regular graph - a graph in which all nodes imitate the same configuration of links - can ever be stable. We finally provide some initial results about the dynamics of non-stable uniform graphs, as individual nodes keep changing their links to improve their cost.

### 4.1 Nash equilibria

The main result of this section is the following.

**Theorem 4.** *For any $n \geq 2$ and any positive integer $k$, uniform stable $(n,k)$-graphs exist, and in any stable graph the cost of any node is $\Theta(1)$ times the cost of any other node. The price of anarchy is $\Omega(\frac{\sqrt{(n/k)}}{\log_k n})$, $O(\sqrt{\frac{n}{\log_k n}})$ (for $k \geq 2$). The price of stability is $\Theta(1)$.*

To prove Theorem 4, we first show fairness. Then we describe a class of stable graphs for any $k$ and prove that they are stable. The graphs in this class have total cost ranging from $O(n^2 \log_k n)$ to $\Omega(n^2 \sqrt{\frac{n}{k}})$. This gives a lower bound on the price of anarchy and the price of stability. Then, we give an upper bound on the diameter of any stable graph and use this to obtain an upper bound on the price of anarchy.

**Lemma 1.** *Fairness: In any stable graph for the $(n,k)$-uniform game, the cost of any node is at most $n + n\lfloor \log_k n \rfloor$ more than, and at most $2 + 1/k + o(1)$ times, the cost of any other node.*

*Proof.* Let $G$ be a stable graph for the $(n, k)$-uniform game and let $r$ be a node in $G$ that has the smallest cost $C^*$. Consider the shortest path tree $T$ rooted at $r$. Let $v$ be any other node. Within $\lfloor \log_k n \rfloor$ hops from $v$, there exists a node $u$ that has at least one edge not in $T$. Since $G$ is stable, node $u$ has cost at most $C^* + n$, since it can achieve this cost by attaching one of its links not in $T$ to $r$. Therefore, the cost of $v$ is at most $C^* + n + n \log_k n$, since the distance from $v$ to any node $w$ is at most $\log_k n$ more than that of $u$ to $w$. Noting that $C^*$ is at least $\sum_{0 \leq i < \log_k n} i k^i \geq (n - n/k) \lfloor \log_k n \rfloor$ completes the proof of the lemma. □

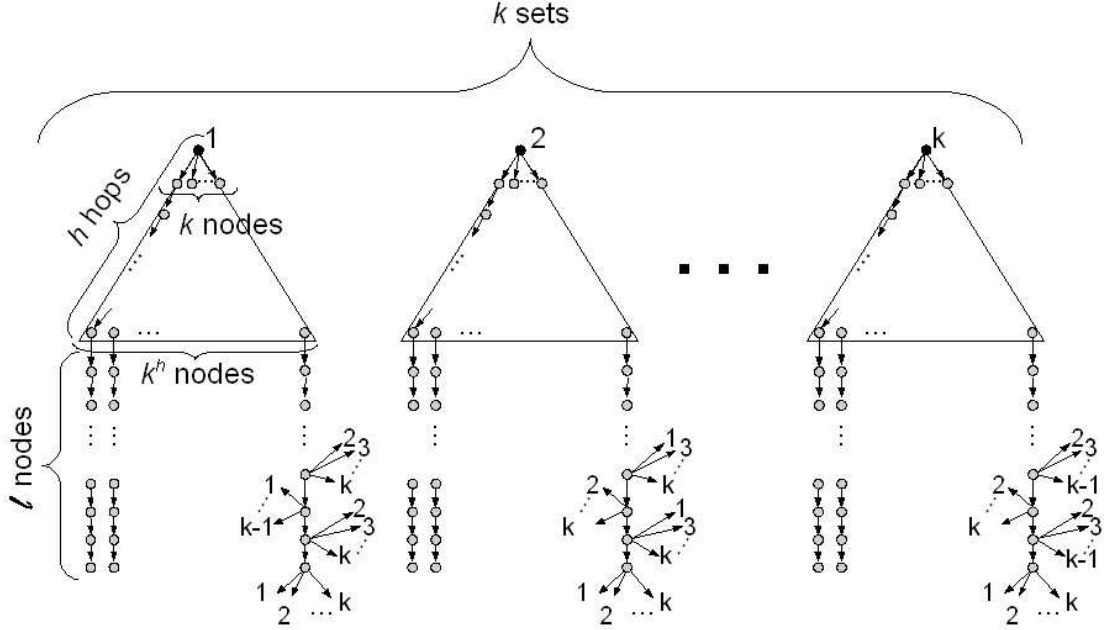

Figure 3: The "Forest of Willows" stable graphs: $k$ sections, each has a complete $k$-ary tree of height $h$. Under each leaf, there is a tail of length $l$. The last node in each tail has an edge to the root of each tree. The second to last node of a tail has an edge to the root of each tree other than its own. The rest of the tail nodes alternate between pointing to all the roots except their own or all the roots except one (arbitrary, but not its own).

In order to give an upper bound on the price of anarchy and the price of stability, we define a class of graphs that is stable. We call this class the "Forest of Willows" graphs (see Figure 3).

**Definition 1.** *Forest of Willows graphs: There are $k$ directed, complete, $k$-ary trees of height $h$ (rooted at nodes $r_1, r_2, \ldots, r_k$). Each of these trees has $k^h$ leaves. Beneath each leaf, there is a tail of length $l$ ($l$ nodes not including the leaf). Let $R_i$ be the nodes in the tree rooted at $r_i$ plus the tails beneath this tree. The last node in each tail has an edge to the root of each of the $k$ trees. The second to last node of a tail in $R_i$ has an edge to each $r_j$, $j \neq i$. If a tail node in $R_i$ does not have an edge to $r_i$, the node above it has an edge to $r_i$ and any $k - 2$ other roots. If a tail node in $R_i$ does have an edge to $r_i$, the node above it has an edge to each $r_j$, $j \neq i$. We call this the* initial configuration. *This graph has $n$ nodes, where $n = k * (2^{h+1} - 1 + 2^h l)$. This can be extended to other values of $n$ by adding additional leaves as evenly as possible across the trees. However, for the sake of simplicity, the following proof of stability assumes that $n$ is of the above form.*

We restrict $h$ and $l$ by requiring:
$$\frac{(h+l)^2}{4} + h + 2l + 1 < \frac{n}{k}$$

By definition of the graph structure, $h \in O(\log_k n)$. Any $l (0 \leq l < 2\sqrt{\frac{n}{k}})$ obey the requirements. Notice that $l < 2\sqrt{\frac{n}{k}}$ implies $h > \frac{\log_2 n}{2} - \frac{\log_2 k}{2} - 1$. Also notice that the diameter of this graph is $\Theta(h + l)$, so as $k$ approaches $\frac{n}{\log^2 n}$, this class converges to a single graph: a collection of $k$ complete $k$-ary trees with edges from the leaves to the roots.

For ease of notation, we use *descendants of $x$* for a node $x \in R_i$ to refer to $x$ plus all nodes $y \in R_i$ such that $x$ is on the unique shortest path from $r_i$ to $y$. We use $D_x$ to refer to the number of descendants of $x$. *Ancestors of $x$* for $x \in R_i$ refer to all the nodes in the shortest path from $r_i$ to $x$ (not including $x$). We use $\delta_x$ to refer to the number of ancestors of $x$ (which is the same as the number of hops from $r_i$ to $x$. When $x$ is clear from the context, we use $D$ and $\delta$ instead of $D_x$ and $\delta_x$.

Since any node that is $\delta$ hops below some $r_i$ is symmetric to any other node $\delta$ hops below any $r_j$, we only need to consider whether nodes in a single $R_i$ (say $R_1$) would move any edges. None of the edges that make up the trees or the tails will be moved, or else the graph would become disconnected. So we only need to consider edges from leaf nodes or tail nodes to roots (call these *non-essential edges*).

With this symmetry in mind, we must verify that no node in $R_1$ will move any of its links. First, we show for that any node $u$ in $R_j$, the number of hops from $r_j$ to $u$ times the number of decendants of $u$ is smaller than the number of nodes in $R_j$ that are *not* descendants of $u$. Intuitively, this is like isolating a single potential link end point: if a node were to move one of its links from $r_j$ to $u$, the decrease to its cost would be smaller than the increase to its cost, even if the distance to each node only increased by one hop. Next, we show that a node would never move its links to one of its own ancestors or descendants, and a node would never place multiple links that have an ancestor/descendant relationship to each other. Once we've eliminated the possibility of related links, it is a relatively small step using our initial lemma to show that no node would ever place its links on non-root nodes. Finally, we show that the nodes would not move their links between roots, completing the proof that Forest of Willows graphs are stable.

The following lemma is used throughout this proof.

**Lemma 2.** *Let $u$ be a given node in $R_1$. If $\delta_u > 1$, then $\frac{n}{k} - D_u - l \geq D_u \delta_u$. If $\delta_u = 1$, then $\frac{n}{k} - D_u \geq D_u$.*

*Proof.* Case 1: $u$ is a tree node (so $1 \leq \delta \leq h$). Here, regardless of the values of $h$ and $l$:
if $\delta > 1$:

$$D = \frac{n}{k^{\delta+1}} - \sum_{i=1}^{\delta} \frac{1}{k^i} < \frac{n}{k^{\delta+1}}$$

$$\frac{n}{k} - D(\delta + 1) - l > n(\frac{1}{k} - \frac{(\delta+1)}{k^{\delta+1}}) - l \geq \frac{n}{k^2} - l \text{ if } k \geq 3, \text{ since } \delta \geq 1$$
$$> 0 \text{ since there are } k \text{ sections with at least } k \text{ tails per section.}$$

and

$$\frac{n}{k} - D(\delta + 1) - l > n(\frac{1}{k} - \frac{(\delta+1)}{k^{\delta+1}}) - l \geq \frac{n}{8} - l > 0 \text{ since } h \geq 3$$

if $\delta = 1$:

$$D = \frac{n}{k^2} - \frac{1}{k}$$

$$\frac{n}{k} - 2D = \frac{n}{k} - \frac{2n}{k^2} + \frac{2}{k} = \frac{n(1 - \frac{2}{k}) + 2}{k} > 0$$

Case 2: $u$ is a tail node.

$$D = h + l - \delta + 1$$
$$\frac{n}{k} - D(\delta + 1) - l = \frac{n}{k} + \delta^2 - (h + l)(\delta + 1) - l - 1$$

The second derivative with respect to $\delta$ is positive, so we only need to check this at the point where $\frac{d}{d\delta} = 0$ (a minima).

$$\delta = \frac{h}{2} + \frac{l}{2} \quad D = \frac{h}{2} + \frac{l}{2} + 1$$

$$\frac{n}{k} - D(\delta + 1) - l = \frac{n}{k} - \frac{(h+l)^2}{4} - h - 2l - 1$$
$$> 0 \text{ by our restrictions on } h \text{ and } l$$

$\square$

**Lemma 3.** *If node $x \in R_1$ benefits by moving any of its non-essential edges to one of its descendants, and if $u_1$ is the closest such descendant, then $x$ will also benefit by moving this edge to another node (distinct from $u_1$) that is $\delta_{u_1}$ hops from a root.*

*Proof.* Suppose $x$ placed at least one of its non-essential edges at node $u_1$, a descendant of $x$. Suppose the $k-2$ other non-essential edges were placed at nodes $u_2, u_3, \ldots u_{k-1}$, and if any other $u_j$ is also a descendant of $x$, then $\delta_{u_j} > \delta_{u_1}$.

The total decrease in hop count by moving the edges from our original placement is at most $\sum_{j=1}^{k-2}(D_{u_j}\delta_{u_j}) - D_{u_1}\delta_x$ (since the sum counts all of the descendants of $u_1$ as having a decrease of $\delta_{u_1}$, but they actually only decreased by $\delta_{u_1} - \delta_x$).

The total increase in hop count is at least $\frac{(k-1)n}{k} - \sum_{j=2}^{k-1}(D_{u_j}) - D_x$ (since each of these non-essential edges used to point to a root, and the distance to all the descendants of these roots that are not also descendants of $x$ or one of the $u_j$ will now increase by at least one hop.)

By moving to another node $\delta_{u_1}$ hops below a root (that is not an ancestor or descendant of $x$ or of any of the other $u_j$), the total decrease in hop count will increase by at least $D_{u_1}\delta_x$. Meanwhile, the increase in hop count can only get lower. Therefore, if $x$ would make the previous move, $x$ would also make the new move. $\square$

**Lemma 4.** *If node $x \in R_1$ benefits by moving any of its non-essential edges to one of its ancestors, $u_1$ ($u_1 \neq r_1$), then $x$ will also benefit by moving this edge to another node $\delta_{u_1}$ hops from a root.*

*Proof.* Suppose $x$ placed at least one of its non-essential edges at node $u_1$, an ancestor of $x$. Suppose the $t-1$ other non-essential edges were placed at nodes $u_2, u_3, \ldots u_t$ ($t$ may be $k-1$ or $k$, depending on the location of $x$). By Lemma 3, we can assume none of $u_2, \ldots u_t$ is a descendant of $x$.

The total decrease in hop count by moving the edges from our original placement is at most $\sum_{j=1}^{t}(D_{u_j}\delta_{u_j}) - D_x\delta_{u_1}$ (since the sum counts all of the descendants of $u_1$ as having a decrease of $\delta_{u_1}$, but actually the descendants of $x$ did not decrease at all).

The total increase in hop count is at least $\frac{tn}{k} - \sum_{j=1}^{t}(D_{u_j})$ (since each of these non-essential edges used to point to a root, and the distance to all descendants of these roots that are not also descendants of one of the $u_j$ will now increase by at least one hop.)

By moving to another node $\delta_{u_1}$ hops below a root (that is not an ancestor or descendant of $x$ or of any of the other $u_j$), the total decrease in hop count will increase by at least $D_x\delta_{u_1}$. Meanwhile, the increase in hop count can only get lower. Therefore, if $x$ would make the previous move, $x$ would also make the new move. $\square$

**Lemma 5.** *If $x$ will benefit by moving any two of its non-essential edges to nodes, $\{u_1, u_2\} \in R_1$, such that $u_1$ is an ancestor of $u_2$, then it will also benefit by moving to a node $\delta_{u_1}$ hops below $r_1$ and a node $\delta_{u_2}$ hops below $r_1$ (neither of which is an ancestor or descendant of $x$ or of any other $u_j$).*

Notice there will always exist two such nodes because there are $k$ branches of each tree and at most $k$ non-essential edges, and $u_2$ must be at least 2 hops below a root, where there are $k^2$ branches (so we can always avoid an ancestor or descendant of $x$ as well).

*Proof.* Suppose $x$ placed two of its non-essential edges at nodes $\{u_1, u_2\}$ such that $u_1$ is an ancestor of $u_2$. Suppose the other non-essential edges were placed at nodes $u_3, u_4, \ldots u_t$ (none of which is an ancestor or descendant of $x$). Also assume there is no $u_j$ on the shortest path from $u_1$ to $u_2$.

The total decrease in hop count by moving the edges from our original placement is at most $\sum_{j=1}^{t}(D_{u_j}\delta_{u_j}) - D_{u_2}\delta_{u_1}$ (since the sum counts all of the descendants of $u_2$ as having a decrease of $\delta_{u_2}$, but actually the decrease was only $\delta_{u_2} - \delta_{u_1}$).

The total increase in hop count is at least $\frac{tn}{k} - \sum_{j=1}^{t}(D_{u_j}) + D_{u_2} - D_x$ (since each of these non-essential edges used to point to a root, and the distance to all nodes that are descendants of these roots but not of $x$ or one of the $u_j$ will now increase by at least one hop.)

By changing the move as suggested in this lemma, the total decrease in hop count will increase by at least $D_{u_2}\delta_{u_1}$. Meanwhile, the increase in hop count can only get lower. Therefore, if $x$ would make the previous move, $x$ would also make the new move. □

**Lemma 6.** *Forest of Willows graphs are stable.*

*Proof.* Consider any possible selections of non-essential edges for a node $x \in R_1$. Suppose $t$ of these, $\{u_1, u_2, \ldots, u_t\}$, are moved away from the roots they point to in the initial configuration (to nodes at least one hop below a root). Also assume that no $u_i$ is an ancestor or descendant of $x$ or of any other $u_j$ (we can make this assumption because of Lemma 3, Lemma 4, and Lemma 5). Then, some nodes in each of $t$ trees will get at least one hop further away from $x$. $D_{u_i}$ nodes will get $\delta_{u_i}$ hops closer (for all $u_i$). $D_x(\leq l)$ nodes will stay the same distance. The change in total hop count is at least the total increase minus the total decrease.

$$\begin{aligned}
\text{change in total hop count} &\geq \frac{nt}{k} - \sum_{i=1}^{t} D_{u_i} - l - \sum_{i=1}^{t} D_{u_i}\delta_{u_i} \\
&= \frac{nt}{k} - \sum_{i=1}^{t} D_{u_i}(\delta_{u_i} + 1) - l \\
&= \sum_{i=1}^{t} \left(\frac{n}{k} - D_{u_i}(\delta_{u_i} + 1)\right) - l \\
&\geq 0 \text{ if } \exists i \text{ such that } \delta_{u_i} > 1, \text{ by Lemma 2}
\end{aligned}$$

When $\delta_{u_i} = 1$ for all $i$, we must consider two cases.

Case 1: $x \in R_1$ does not have an edge to $r_1$ in the initial configuration (or does not move this edge). In this case, the total increase is at least $\frac{nt}{k} - \sum_{i=1}^{t} D_{u_i}$. (The $-l$ is not there, because $x$ is not located under a root that increases.) This gives a change in total hop count $\geq \frac{nt}{k} - \sum_{i=1}^{t} D_{u_i}(\delta_{u_i} + 1) > 0$ (by the $\delta = 1$ condition in Lemma 2).

Case 2: $x \in R_1$ has an edge to $r_1$ in the initial configuration and moves this edge. All of the nodes $u_i$ are 1 hop below roots, and none is an ancestor of $x$. There is a single node, $u_1$, that is 1 hop from $r_1$ that is an ancestor of $x$: the distance to each of the descendants of $u_1$ that are not also descendants of $x$ (at least $D_{u_1} + 1 - l$ nodes) will increase by at least 2 hops ($x$ cannot be the second to last node in a tail because it had an edge to $r_1$. If $x$ is the last node of a tail, then the new distance to $r_1$ is at least $h - 1$. If $x$ is at least 2 hops from the end of a tail, then there are at least 2 hops to the closest node pointing to $r_1$).

Therefore, the total increase in trees other than $R_1$ is at least $\frac{n(t-1)}{k} - \sum_{i=1}^{t-1} D_{u_i}$, and the increase in $R_1$ is at least $2(D_{u_1} + 1 - l)$. This gives the following total change in hop count.

$$\begin{aligned}
\text{change in total hop count} &\geq \frac{n(t-1)}{k} - \sum_{i=1}^{t-1} D_{u_i}(\delta_{u_i} + 1) + (2 - \delta_{u_1})D_{u_1} + 2 - 2l \\
&= \sum_{i=1}^{t-1} \left(\frac{n}{k} - 2D_{u_i}\right) + D_{u_1} + 2 - 2l \\
&\geq 0 \text{ by Lemma 2 and the fact that } D_{u_1} \text{ includes at} \\
&\quad \text{least 3 tails (when } \delta_{u_1} = 1\text{) as long as } k > 1 \text{ and } h \geq 3.
\end{aligned}$$

Therefore, $x$ does not have incentive to move any of its non-essential edges to nodes other than roots.
Finally, we must verify that $x$ has no incentive to move an edge from one root to another.
Case 1: $x \in R_1$ has edges to all roots except $r_1$ in the initial configuration.

In this case, consider what would happen if $x$ moved an edge from some root $r_j$ to $r_1$. The distance to descendants of $r_1$ but not of $x$ would decrease by at most 1 hop, since the node beneath $x$ in the tail already has an edge to $r_1$. This is a decrease of at most $\frac{n}{k} - 2$ ($x$ has only $k-1$ non-essential edges, so at least $x$ and one node below it keep the same distance). Meanwhile, the distance to all the descendants of $r_j$ will increase by at least one hop. This gives an increase of at least $\frac{n}{k}$. Since the increase is always larger than the decrease, there is no incentive for this move.

Case 2: $x \in R_1$ has edges to all roots except some $r_j$ ($j \neq 1$) in the initial configuration.

In this case, first consider what would happen if $x$ moved an edge from $r_1$ to $r_j$. The distance to descendants of $r_1$ but not of $x$ (at least $\frac{n}{k} - l$ nodes) would increase by at least 2, since it is 2 hops to another node with an edge to $r_1$. So there is an increase of at least $\frac{2n}{k} - 2l$. The distance to the $\frac{n}{k}$ descendants of $r_j$ would decrease by 1 hop, since the node beneath $x$ already points to $r_j$. So the decrease is at most $\frac{n}{k}$. The increase is always larger than the decrease, so there is no incentive for this move.

Next consider what would happen if $x$ moved an edge from some $r_g \neq r_1$ to $r_j$. The distance to the $\frac{n}{k}$ descendants of $r_g$ would increase by 1 hop, while the distance to the $\frac{n}{k}$ descendants of $r_j$ would decrease by 1 hop. Therefore, this move does not make any difference to $x$, so $x$ has no incentive to move. □

**Lemma 7.** *The diameter of any uniform stable $(n,k)$-graph ($k \geq 2$) is $O(\sqrt{n \log_k n})$, and there is at least one node whose distance to any other node is $O(\sqrt{n})$.*

*Proof.* Let $G$ be a stable graph for the $(n,k)$-uniform game, and let $\Delta$ denote the diameter of $G$, given by a path from a node $r$ to a node $v$. Consider a shortest path tree from $r$; so the depth of this tree is $\Delta$ and $v$ is a leaf of $T$. Let $P$ denote the set of nodes on the path from $r$ to $v$ in $T$, not counting $r$; so $|P| = \Delta$. Let $C$ be the sum of distances from $r$ to the $n - \Delta$ nodes not in $P$. The sum of distances from $r$ to the $\Delta$ nodes in $P$ is exactly $\Delta(\Delta + 1)/2$. So the cost of $r$ is $C + \Delta(\Delta + 1)/2$.

The cost of $v$ is at most $C + n - \Delta/2 + \Delta(\Delta/2 + 1)/4 + \Delta(\Delta/2 + 1)/4$ since $v$ can use one of its at least two edges to connect to $r$ and the other to connect to a node halfway along the path from $r$ to $v$. Simplifying, we obtain that the cost of $v$ is at most $C + n + \Delta^2/4$. By Lemma 1, the cost of $v$ is at least $C + \Delta(\Delta + 1)/2 - n - n \log_k n$. We thus obtain the inequality:

$$C + n + \Delta^2/4 \geq C + \Delta(\Delta + 1)/2 - n - n \log_k n,$$

yielding $\Delta = O(\sqrt{n \log_k n + 2n})$.

Using the fact that the cost of $v$ is at least $C$ (in place of the reference to Lemma 1) in the above proof gives the second part of the lemma. □

*Proof of Theorem 4:* The first claim directly follows from Lemma 1. In any graph with max degree $k$, every node must have cost at least $\Omega(n \log_k n)$. Forest of Willows graphs with $l = 0$ have total cost per node $O(n \log_k n)$. Therefore, the price of stability is $\Theta(1)$.

If $l = 0$, a Forest of Willows graph has total cost per node $= O(n \log_k n)$. Therefore, the social utility has total cost (over all nodes) $O(n^2 \log_k n)$. If $l = \Omega(\sqrt{\frac{n}{k}})$, the total cost (over all nodes) is $\Omega(n^2 \sqrt{\frac{n}{k}})$. Therefore, the price of anarchy is $\Omega(\frac{\sqrt{(n/k)}}{\log_k n})$.

Finally, Lemma 7 implies that the total cost of any node in the worst Nash equilibium cannot be higher than $O(\sqrt{n \log_k n})$, so the total cost is $O(n\sqrt{n \log_k n})$. We already know that the social equilibrium is at least $O(n \log_k n)$. Therefore, the price of anarchy is $O(\sqrt{\frac{n}{\log_k n}})$. $\square$

## 4.2 Stability of regular graphs

In the context of overlay or peer-to-peer networks, a natural degree-$k$ graph to consider is to map the nodes to $\mathbb{Z}_n = \{0, 1, \ldots, n-1\}$ and have the $k$ edges for all nodes be defined by $k$ offsets $a_i$, $0 \le i < k$: the $i$th edge from node $x$ goes to $x + a_i$ mod $n$. We refer to such graphs as *regular graphs*. For a suitable choice of the offsets, these graphs have diameter $O(n^{1/k})$. In this section, we study Abelian Cayley graphs, a more general class that includes regular graphs. We show that these graphs are not stable for $k \ge 2$. Cayley graphs have been widely studied by mathematicians and computer scientists, and arise in several applications including expanders and interconnection networks (e.g., see [2, 3, 8]).

A Cayley graph $G(H, S)$ is defined by a group $H$ and a subset $S$ of $k$ elements of $H$. The elements of $H$ form the nodes in $G$, and we have an edge $(u, v)$ in $G$ if and only if there exists an element $a$ in $S$ such that $u \cdot a = v$, where $\cdot$ is the group operation. A Cayley graph $G(H, S)$ is referred to as an Abelian Cayley graph if $H$ is Abelian (that is, the operation $\cdot$ is commutative). The regular graph described in the preceding paragraph is exactly the Cayley graph with the group $H$ being the Abelian additive group $\mathbb{Z}_n$ and $S = \{a_i \bmod n : 0 \le i \le k\}$.

We prove that no pure Nash equilibria exist in Abelian Cayley graphs, using a particular embedding of these graphs into $k$-dimensional grids. Let $G(H, S)$ be a given Abelian Cayley graph and let the $k$ elements of $S$ be $a_i$, $0 \le i < k$. We assume without loss of generality that $S$ does not contain the identity of $H$ since these edges only form self-loops, which clearly cannot belong to any stable graph. Each edge of the graph $G$ can be labeled by the index of the element of $S$ that creates it; that is, if $v = u \cdot a_i$, then we call the edge $(u, v)$ an *i-edge*. The edge labels naturally induce labels on paths as follows. If a path contains $x_i$ $i$-edges, then we label the path by the vector $\vec{x} = (x_1, \ldots, x_i, \ldots, x_k)$. Note that the length of a path with label $\vec{x}$ is simply $\sum_{1 \le i \le k} x_i$. Furthermore, the commutativity of the underlying group operator implies that for all nodes $v$ and all path labels $\vec{x}$, every path that starts from $v$ and has label $\vec{x}$ ends at the same node.

We say that node $v$ has label $\vec{x}$ if there exists a shortest path from $r$ to $v$ that has label $\vec{x}$. For any node $v$, while two shortest paths from $r$ to $v$ share the same sum of label-coordinates, the actual path labels may be different; therefore, a node may have multiple labels. However, a particular label is assigned to at most one node.

We next prove that for $k \ge 2$ no Abelian Cayley graph is stable. For $k = 1$, it is trivial to see that the simple directed cycle is an Abelian Cayley graph and is stable.

**Theorem 5.** *For any $k \ge 2$, no Abelian Cayley graph with degree $k$ and $n$ nodes is stable, for $n \ge c2^k$, for a suitably large constant $c$.*

*Proof.* We now consider the impact of replacing the $i$-edge from root $r$ to $r_i = r \cdot a_i$ by the edge from $r$ to $r'_i = r \cdot a_i \cdot a_i$. The node $r$ equals $(0, 0, \ldots, 0)$, while the node $r_i$ equals $(0, 0, \ldots, 1, \ldots, 0)$ with a 1 in the $i$th coordinate. (We note that $r$ and $r_i$ are distinct since $a_i$ is not identity.) For every node $v$ that has a label $\vec{v}$ such that $v_i \ge 2$, the distance decreases by 1. Let $S_i = \{v : v \text{ has a label } \vec{v} \text{ with } v_i \ge 2\}$ be the set of such nodes. On the other hand, the only node whose distance from $r$ increases is the node $r_i$; this is because any path in the original graph starting from $r$, having exactly one $i$-edge $(r, r_i)$ and having length at least two, can be substituted by another path of the same length with an $i$-edge as its second edge.

We bound the increase in the distance from $r$ to $r_i$ in terms of the diameter $\Delta$ of the graph. Let $w = r \cdot a_j^{-1} \ne r_i$ denote a node that has an edge to $r$ in $G$. Since the shortest path to any vertex other than $r_i$ has not increased, the distance from $r$ to $r_i$ is at most $\Delta + 2$, given by a shortest path from $r$ to $w$, followed by an $i$-edge and then by a $j$-edge. Thus, when the edge $(r, r_i)$ is replaced by the edge $(r, r_i')$, the total utility for node $r$ decreases by at least $|S_i| - (\Delta + 2)$. By the definition of $S_i$, this is precisely the set

$$G \setminus \bigcup_{0 \leq i < k} S_i = \{\vec{v} : 0 \leq v_i \leq 1 \text{ for all } i\}.$$

Therefore, there exists $i$, $0 \leq i < k$, such that $|S_i| \geq (n - 2^k)/k$, and the graph $G$ is not stable if $(n - 2^k)/k$ exceeds $\Delta + 1$. By Lemma 7, for $G$ to be stable we must have $\Delta = O(\sqrt{n \log_k n})$. Using this upper bound on $\Delta$, if $n \geq c2^k$ (for an appropriately large constant $c$), then $(n - 2^k)/k$ exceeds $\Delta + 1$, implying that $G$ is not stable. □

**Corollary 1.** *For any $k > 4$, the $2^k$-node hypercube is not stable for the $(2^k, k)$-uniform game.*

If the degree $k$ is more than nearly half the size of the graph, then any degree-$k$ $n$-node Abelian Cayley graph is stable.

**Lemma 8.** *For all $k > \frac{n-2}{2}$ any degree-$k$ $n$-node Abelian Cayley graph is stable.*

## 4.3 Dynamics of best response walks

Given the existence of pure Nash equilibria for $(n, k)$-uniform games, it is natural to ask whether an equilibrium can be obtained by a sequence of local links changes. In particular, we consider a specific type of best response walk: in each step, a node tests for its stability and, if it is not stable, moves its links to the set of nodes that optimize its cost. We assume for convenience that only one node attempts to change its links in any step of the best response walk.

We first show that, starting from any initial state, the best response walk converges to a strongly connected graph in $O(n^2)$ steps, as long as every node is allowed to execute a best response step once every $n$ steps. Furthermore, there exists an initial state such that a best response walk takes $\Omega(n^2)$ steps to converge to strong connectivity. We next study convergence to stability and show that there exists an initial state from which a particular best response walk does not converge to a stable graph. This means that the $(n, k)$-uniform game is *not an ordinal potential game*, a characteristic which justifies our use of a constructive proof for the existance of Nash equilibria.

**Convergence to a strongly connected graph**. For a given node $u$, we define the *reach* of $u$ to be the number of nodes to which it has paths. Since the cost of disconnection is assumed to be $M > n$, when we execute a best-response step for a node $u$, the reach of $u$ cannot decrease.

**Lemma 9.** *Suppose the graph $G$ is not strongly connected, and a node $u$ changes its edges according to a best response step. Then, after the step, the reach of any node other than $u$ either remains the same or is at least the new reach of $u$.*

*Proof.* If a node $v$ has a path to $u$, then the reach of $v$ is at least the reach of $u$ after the best response step. Otherwise, the reach of $v$ does not change. □

The above lemma indicates that whenever a best response step causes a change, the vector that consists of all the reach values in increasing order becomes lexicographically larger. In order to show convergence, we need to argue progress. We will do so by showing that whenever the graph is not strongly connected, there exists a node that can improve its reach. In fact, we use a stronger property that allows us to bound the convergence time.

Consider best response walks that operate in a round-robin manner. In each round, each node (one at a time in an arbitrary order) executes a best response step. The order may vary from round to round. Let $G_r$ refer to the graph before round $r$.

**Lemma 10.** *If $G_r$ is not strongly connected at the start of round $r$, then the minimum reach increases by at least one during the round.*

*Proof.* Consider the strongly connected components of the given graph $G_r$. Consider the component graph $CG$ in which we have a vertex for each strongly connected component and edge between two components whenever there is an edge from a vertex in one component to the other. This graph is a dag. Let $m$ denote the minimum reach in $G_r$. By Lemma 9, nodes with reach greater than $m$ will continue to have reach greater than $m$. So we only need to consider nodes with reach $m$. All of these nodes lie in sink components.

Consider any sink component $C$. We first argue that there exists a node in $C$ that can improve its reach by executing a best response step. Consider a vertex $u$ in $C$ that has an edge from a vertex $v$ in another component. Let $w$ be a vertex in the sink component that has an edge to $u$. All of $u$, $v$, and $w$ exist by definition of strongly connected components (and our assumption that the out-degree of every vertex is at least 1). If $w$ replaces the edge $(w, u)$ with $(w, v)$, it can reach all vertices in the sink component as well as the component containing $v$. The latter set is clear; for the former set, note that all we have done is "replace" the direct edge $(w, u)$ by the two-hop path $w \to u \to v$.

For any sink component $C$, let $v$ be the first node in $C$ in the round order that improves its reach through a best response step. Note that $v$ exists by the argument of the preceding paragraph. Furthermore, in the step prior to $v$'s best response, the reach of every node in $C$ is $m$. After $v$'s best response, the reach of $v$ increases to at least $m+1$, as does that of *every* node in $C$, since they each have a path to $v$. By Lemma 9, after every subsequent step, the reach of any node in $C$ is at least $m+1$. Therefore, it follows that at the end of the round, the reach of every node in a sink component of $CG$ increases; hence, the minimum reach increases, completing the proof of the lemma. □

**Theorem 6.** *The best response walk converges to a strongly connected graph in $n^2$ steps.*

*Proof.* By Lemma 10, the minimum reach increases by at least one. Since the initial reach is 1 and the maximum reach is $n$, the number of steps for the best response walk to converge to a strongly connected graph is at most $n^2$. □

The above theorem is essentially tight. In the following scenario (with $k = 1$), a best response walk may take $\Omega(n^2)$ steps to converge to a strongly connected graph. Consider a graph $G$ of $n = r + p$ nodes that is a directed ring over $r \geq n/2$ nodes together with a directed path of $p = n - r$ nodes that ends at one of the nodes in the ring. Suppose a round begins at the tail $T$ of the directed path, which can reach all nodes, proceeds along the path and then along the ring in the direction of the ring. The $p$ nodes on the path cannot improve their reach. Furthermore, the first $r - p$ nodes on the ring (in round-robin order) also cannot improve their reach in a best response step. The $(r - p + 1)$st node can improve its reach by connecting to $T$, yielding a new graph $G'$ that is a directed ring over $r + 1$ nodes and a directed path of $n - r$ nodes. If we repeat this process, the number of steps to converge is $\Omega(n^2)$.

**Cycles in best response walks**. Unlike strong connectivity, convergence to a pure Nash equilibrium is not guaranteed. In the following simple example, a round-robin best-response walk contains loops. This simple example is a (7,2)-uniform game that starts from the top-left configuration of Figure 4. The nodes take turns in round-robin order, starting with node 6 then nodes 0,1,2, and so on. Tracing the example, one can verify that after 6 deviations (nodes $6, 3, 2, 6, 3, 2$ re-linking in this order, implying that missing nodes are stable), the graph returns to the initial configuration thus completing a loop.

The above example of a loop in the best response walk shows that the uniform-$(n, k)$-game is not an ordinate potential game. However, the loop does not rule out the possibility that either (a) a well-chosen best response walk converges from any initial state, or (b) certain best response walks do converge to stability if started from simple initial configurations such as the empty graph.

We have observed experimentally that best response walks in which a node with the maximum cost always makes the next best response step does not always converge to a stable graph. However, based on our experimental data, this best response walk starting from an empty graph does seem to converge to a stable graph. Our experiments also suggest that there may be some exponentially long best-response paths that start in some non-empty initial configuration and end at a stable graph.

# 5 Max distance utility function

In the BBC games we have studied thus far, the utility of a node $u$ in $G(S)$ given by $-\sum_v w(u,v)d(u,v)$, where $d(u,v)$ is the shortest path from $u$ to $v$ in $G(S)$ according to the lengths given by $\ell$. We have also considered a natural variant of the utility function: the utility of $u$ is $-\max_v w(u,v)d(u,v)$. In order to make it clear that we are using a different cost function, we will call the max distance version *BBC-max games*.

As with the previous cost function, we show there exist instances of the general BBC-max game that have no Nash equilibrium. If we restrict ourselves to the uniform version (uniform $(n,k)$-BBC-max game), there is a stable graph for any $n$ and $k < n$. It turns out that ratio between the total utility achieved in a Nash equilibrium in a uniform BBC-max game and the social optimum could be much worse than in BBC games. In particular, we establish a lower bound of $\Omega(\frac{n}{k \log_k n})$ on the price of anarchy in BBC-max games.

**Theorem 7.** *For all $n \geq 16$, $k \geq 1$, there exists a nonuniform BBC-max game with $n$ nodes, uniform link costs, uniform link lengths, and a uniform budget of $k$ for every node, such that the game has no pure Nash equilibrium. Furthermore, it is NP-hard to determine whether a given nonuniform BBC-max game instance is a Nash equilibrium.*

*Proof.* Figure 5 depicts a modified version of the gadget of Figure 1, in which the following have been added: 0LT is linked to the (previously not depicted) "sink" node 0S; the sink is linked to two new nodes, the second of which connects to 0C. Similar additions apply to gadget1 but are not depicted. As in Theorem 1, solid lines can be implemented by setting the preference of the source node for the destination node to a positive value and zeroing out its preference for all other nodes. The central "switch" at node 0C can be implemented as in Theorem 1.

The bottom switch, say of node 0RB, can be implemented by setting $w_{0RB,0S} = w_{0RB,0C} = a > 0$, and zeroing out all other preference values for 0RB. This way, if edge (0C,0LT) is not implemented, 0RB will connect to 0S, because this gives it a cost of $a \cdot 4$ (gets 0S with 1 hop, and through it, 0C with 4 hops), which is minimal among all its other options (connecting to 0C would make its cost $a \cdot M$; connecting to 0LT would make it $a \cdot 5$). If now edge (0C,0LT) is implemented, then 0RB will connect to 0C and achieve a minimum cost of $a \cdot 3$ given by the diameter 0RB→0C→0LT→0S (connecting to 0RB would cost $a \cdot 4$ and to 0LT $a \cdot 5$). This assignment of weight gives rise to the looping described in Theorem 1, therefore proving the current theorem. □

**Theorem 8.** *The Price of Anarchy for uniform $(n,k)$ BBC-max games is $\Omega(\frac{n}{k \log_k n})$.*

*Proof.* Consider the following graph for $k > 2$. There are $2k - 1$ tails, $\{t_1, t_2, \ldots, t_{2k-1}\}$, each of length $l = \frac{n-1}{2k-1}$. There is also one "root" node $r$ with edges to the top node in $t_1, t_2, \ldots, t_k$. For ease of notation, we will define segments $S_1 = \{r, t_1, t_2, \ldots, t_k\}$, $S_2 = \{t_{k+1}\}, S_2 = \{t_{k+2}\}, \ldots, S_k = \{t_{2k-1}\}$, with the head of each segment $S_i$ ($i > 1$) = the first node of the tail, and the head of $S_1 = r$. The last node of each tail points to the head of each segment. The rest of the nodes in each tail point to $r$ and to the last node of a tail. The location of the rest of the edges don't matter. See Figure 6. We will show that this graph is a Nash equilibrium.

First consider whether a node at the end of a tail would benefit by moving any of its edges. Its current max distance $= 2 + l$ (to a node at the end of $t_1$, $t_2$, or $t_3$). If it does not have one edge to each segment, then the max distance is at least $2 + l$ (since it takes at least one hop to get to a node that will point to the segment, and all other nodes that point to the segment point to the head). If the one edge pointing to some segment does not point to the head of the segment, the max distance is 1 (to get to the segment) + the distance to the end of the tail + 1 (to get to the head) + the distance back to where it started $= 2 + l$.

Next, consider whether a node in the middle of a tail would benefit by moving any of its edges. Its current max distance $= 2 + l$ (the same distance to the end of any tail other than the tail it lives in). In order to get closer to the end of *every* other tail (since all are currently the same max distance), it would need to point closer to the middle of each tail. For segments $S_2, \ldots S_k$ (other than its own segment), this would be possible by pointing an edge to the head of each segment (or anywhere within the segment). In order to shorten all of these distances, at least $k - 2$ edges must be used. However, the only way to reduce the distance to nodes in $S_1$ would be to point an edge to each of the $k$ tails within the segment (or to $k-1$ edges if this node lives in $S_1$). There are not enough edges to improve distances to $S_1$ and to all other tails. Therefore, this node cannot improve its utility.

This example can be extended to the case where $k = 2$ with a small adjustment. In this case, there are 3 paths plus one node that points to the head of two of the paths. The nodes at the end of each path point to the root of the single path and the extra node. The second to last nodes in the other paths point to the extra node. The rest of the nodes in the other paths point to the end of a tail.

In the Forest of Willows graphs described in Section 4.1, when $l = 0$ the sum of the max distances $= O(n \log_k n)$. Therefore, the social optimum cost is at most $O(n \log_k n)$. We have just shown that there is a graph with the sum of the max distances $= \Omega(\frac{n^2}{k})$. Therefore, the Price of Anarchy is $\Omega(\frac{n}{k \log_k n})$. □

**Theorem 9.** *The Price of Stability for uniform $(n, k)$ BBC-max games is $\Theta(1)$.*

*Proof.* It is easy to verify that the Forest of Willows graphs with $l = 0$ (described in section 4.1) are also stable under the max cost function. Obviously, no node can have max distance less than $\log_k n$. Therefore, the social optimum is at most $O(n \log_k n)$, and the best Nash is at least $\Omega(n \log_k n)$, so Price of Stability is $\Theta(1)$. □

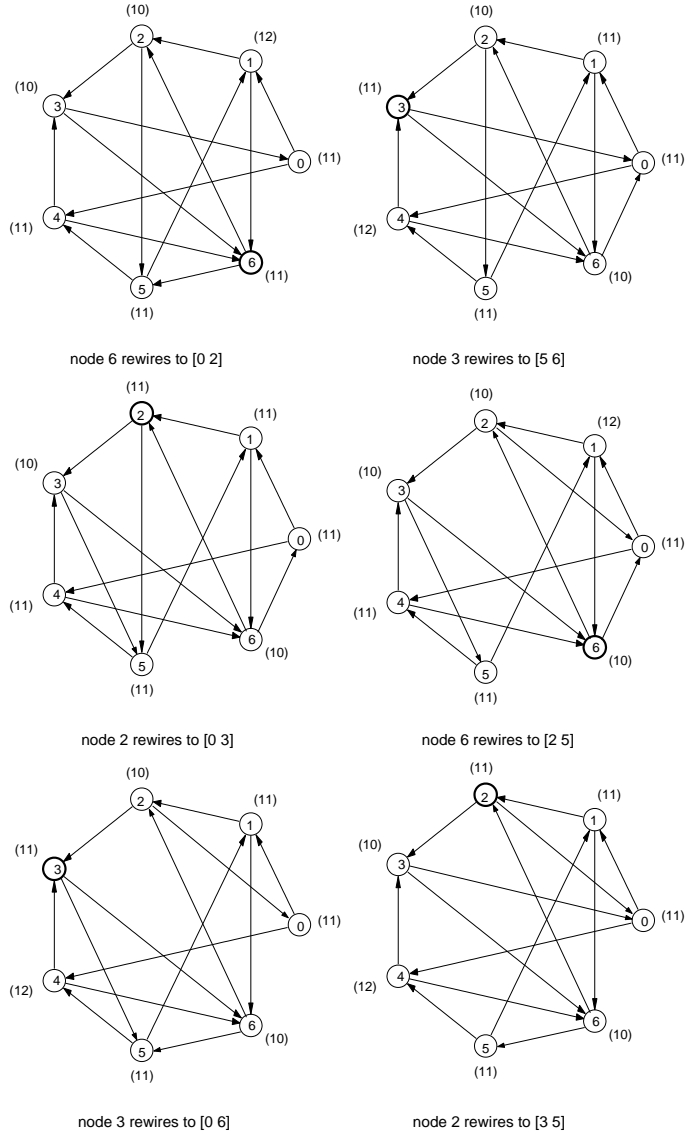

Figure 4: An example in which a round-robin best-response walk loops. Starting from the top left configuration and following a round-robin best-response walk $6 \to 0 \to 1 \to \ldots \to 6 \to 1 \ldots$ we get back to the initial configuration after 6 deviations (nodes $6, 3, 2, 6, 3, 2$). Turns that are not illustrated imply stable nodes. Next to each node we indicate its cost under the current configuration.

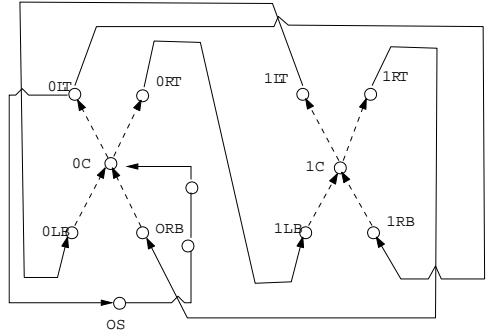

Figure 5: Modified gadget - consisting of two subgadgets

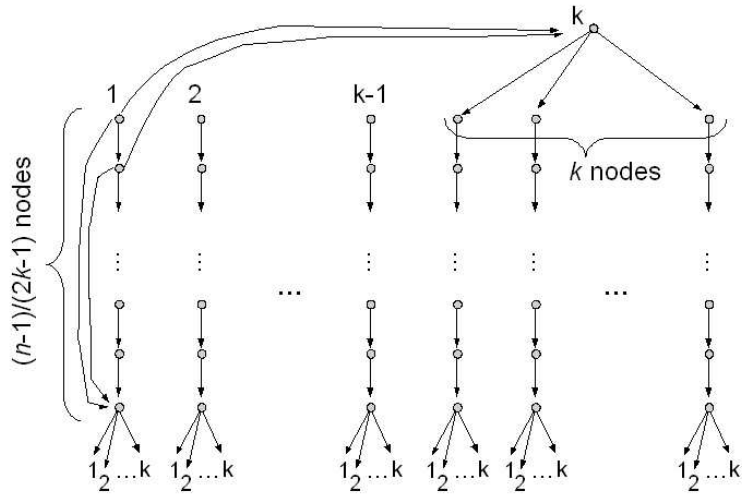

Figure 6: A high cost Nash equilibrium for the max distance cost function: $2k-1$ paths, one node points to $k$ of them. Each node at the end of a path points to the start of the first $k-1$ paths and the extra node. Each node in a path points to the node at the end of the path and to the extra node. The rest of the edge don't matter.

21